\documentclass[12pt]{article}
\usepackage{amsmath}
\def\be{\begin{equation}}
\def\ee{\end{equation}}
\def\arr{\begin{array}{rll}}
\def\ea{\end{array}}
\def\bea{\begin{eqnarray}}
\def\eea{\end{eqnarray}}

\def\N2{$N{=}2$}

\def\>{\rangle}
\def\<{\langle}
\def\+{\dagger}
\def\={\ =\ }

\begin{document}
\renewcommand{\thefootnote}{\fnsymbol{footnote}}
\begin{titlepage}
\setcounter{page}{0}
\vskip 1cm
\begin{center}
{\LARGE\bf Schwarzian mechanics via }\\
\vskip 0.5cm
{\LARGE\bf  nonlinear realizations }\\
\vskip 1.5cm
$
\textrm{\Large Anton Galajinsky\ }
$
\vskip 0.7cm
{\it
Tomsk Polytechnic University, 634050 Tomsk, Lenin Ave. 30, Russia
} \\
\vskip 0.1cm
{E-mail: galajin@tpu.ru}
\vskip 0.5cm

\end{center}
\vskip 1cm
\begin{abstract} \noindent
The method of nonlinear realizations is used to clarify some conceptual and technical issues related to the Schwarzian mechanics.
It is shown that the Schwarzian derivative arises naturally, if one applies the method to $SL(2,R)\times R$ group and decides to keep the number of the independent Goldstone fields to a minimum. The Schwarzian derivative is linked to the invariant Maurer--Cartan one--forms, which make its $SL(2,R)$--invariance manifest.
A Lagrangian formulation for a variant of the Schwarzian mechanics studied recently in [Nucl. Phys. B 936 (2018) 661] is built and its geometric description in terms of $4d$ metric of the ultrahyperbolic signature is given.
\end{abstract}


\end{titlepage}

\renewcommand{\thefootnote}{\arabic{footnote}}
\setcounter{footnote}0

\noindent
{\bf 1. Introduction}\\

\noindent
When first encountering the Schwarzian derivative \cite{HS,OT}
\be\label{SD}
S(\rho(t))=\frac{\dddot{\rho}(t)}{\dot\rho(t)}-\frac 32 {\left(\frac{\ddot{\rho}(t)}{\dot\rho(t)}\right)}^2,
\ee
where $\rho(t)$ is a real function,
one may be amazed by its invariance under the $SL(2,R)$ transformations\footnote{To be more precise, $S(\rho(t))$ holds invariant under the fractional linear transformation $\rho'(t)=\frac{a \rho(t)+b}{c \rho(t)+d}$ with $ad-cb\ne0$. Because the matrices $-\left(\begin{array}{cc}
                                                       a & b \\
                                                       c & d
                                                      \end{array}\right)$ and $\left(\begin{array}{cc}
                                                       a & b \\
                                                       c & d
                                                      \end{array}\right)$
result in the same transformation, the actual symmetry is $GL(2,R)/Z_2$.}
\be\label{tr}
\rho'(t)=\frac{a \rho(t)+b}{c \rho(t)+d}, \qquad ad-cb=1, \qquad \Rightarrow \qquad S(\rho'(t))=S(\rho(t)).
\ee
Because $sl(2,R)$ is a finite--dimensional subalgebra of the Virasoro algebra, the Schwarzian derivative arises naturally within the context of string theory and related field theories (see, e.g., Chapter 4 in Ref. \cite{DF}). In recent years there has been a burst of activity in studying $1d$
quantum mechanics that arises as the low energy limit of the solvable theory displaying maximally chaotic  behaviour -- the so called Sachdev--Ye--Kitaev model.\footnote{The literature on the subject is rather extensive. For an introduction and references to the original literature see, e.g., \cite{MTV}.} A peculiar feature of the system is that its Lagrangian density is proportional to the Schwarzian derivative of a specific function.

As $S(\rho(t))$ in (\ref{SD}) is $SL(2,R)$--invariant, {\it any} function of it can be used to define the equation of motion of a higher derivative $1d$ mechanics enjoying $SL(2,R)$ symmetry. In a recent work \cite{AG}, a variant of the Schwarzian mechanics was studied which was governed by the third order equation of motion $S(\rho(t))=\lambda$, where $\lambda$ is a coupling constant.
It was shown that in general the model undergoes stable evolution but for one fixed point solution which exhibits runaway behavior. Conserved charges associated with the $SL(2,R)$ symmetry have been constructed by integrating the equation of motion and linking constants of integration to $\rho(t)$ and its derivatives. Yet, the Hamiltonian formulation in \cite{AG} was unconventional and a Lagrangian formulation was missing.

The goal of this paper is to apply the method of nonlinear realizations \cite{CWZ} so as to clarify some conceptual and technical issues related to the Schwarzian mechanics.

We begin by demonstrating in Sect. 2 that the Schwarzian derivative arises naturally, if one applies the method to $SL(2,R)\times R$ group and decides to keep the number of the independent Goldstone fields to a minimum. Furthermore, $S(\rho(t))$ is linked to the invariant Maurer--Cartan one--forms, which make its $SL(2,R)$--invariance manifest.

In Sect. 3, the Maurer--Cartan one--forms are used to build a Lagrangian formulation for a variant of the Schwarzian mechanics studied recently in \cite{AG}. The full set of conserved charges is found. Similarities and differences between the Schwarzian mechanics and the conformal mechanics by de Alfaro, Fubini and Furlan \cite{DFF} are discussed.

Sect. 4 is focused on a geometric description of the Schwarzian mechanics in terms of $4d$ metric of the ultrahyperbolic signature which obeys the Einstein equations.

Some final remarks are gathered in the concluding Sect. 5.

\vspace{0.5cm}

\noindent
{\bf 2. Schwarzian derivative via the method of nonlinear realizations}\\

\noindent
In what follows, we will need the infinitesimal form of the $SL(2,R)$ transformation exposed in Eq. (\ref{tr}) above
\be\label{inftr}
\rho'(t)=\rho(t)+\alpha, \qquad \rho'(t)=\rho(t)+\beta \rho(t), \qquad \rho'(t)=\rho(t)+\gamma \rho^2(t).
\ee
The corresponding generators
\be\label{pdk}
P=i \partial_\rho, \qquad D=i \rho \partial_\rho, \qquad K=i \rho^2 \partial_\rho
\ee
are associated with the translation, dilatation, and special conformal transformation acting upon the {\it form} of the field $\rho(t)$.
They obey the structure relations of $SL(2,R)$ algebra
\be\label{sl2r}
[P,D]=i P, \qquad [P,K]=2 i D, \qquad [D,K]=i K.
\ee

One more symmetry operator, which commutes with $(P,D,K)$, is related to the time translation
\be\label{u1}
H=i \partial_t, \quad \Rightarrow \quad t'=t+\sigma, \qquad \rho'(t')=\rho(t).
\ee

Let us demonstrate that the Schwarzian derivative (\ref{SD}) comes about naturally,
if one applies the method of nonlinear realizations \cite{CWZ} to $SL(2,R)\times R$ group and keeps the number of the independent Goldstone fields to a minimum.

As the first step, one considers a space parametrized by the temporal variable $t$ and equipped with the Goldstone fields $\rho(t)$, $s(t)$, $u(t)$, whose generic element reads\footnote{It should be born in mind that both the form of the $SL(2,R)$ transformations and the invariant Maurer--Cartan one--forms essentially depend on the order of the factors entering the right hand side of (\ref{g}). The choice (\ref{g}) proves to be optimal. }
\be\label{g}
\tilde g=e^{itH} e^{i\rho(t)P} e^{i s(t) K} e^{i u(t) D}.
\ee
The left multiplication by a group element $\tilde g'=g \cdot \tilde g$, where
$g=e^{i\sigma H} e^{i \alpha P} e^{i \gamma K} e^{i \beta D}$ and $(\sigma,\alpha,\gamma,\beta)$ are infinitesimal parameters,
defines the action of the group on the space.
Taking into account the Baker--Campbell--Hausdorff formula
\be\label{ser}
e^{iA}~ T~ e^{-iA}=T+\sum_{n=1}^\infty\frac{i^n}{n!}
\underbrace{[A,[A, \dots [A,T] \dots]]}_{n~\rm times},
\ee
one gets
\begin{align}\label{tr1}
&
\rho'(t)=\rho(t)+\alpha, && s'(t)=s(t), && u'(t)=u(t)
\nonumber\\[4pt]
&
\rho'(t)=\rho(t)+\beta \rho(t), && s'(t)=s(t)-\beta s(t), && u'(t)=u(t)+\beta,
\nonumber\\[4pt]
&
\rho'(t)=\rho(t)+\gamma \rho^2(t), &&  s'(t)=s(t)+\gamma(1-2\rho(t) s(t)), &&  u'(t)=u(t)+2\gamma \rho(t),
\end{align}
along with
\be\label{tr2}
t' =t+\sigma, \qquad \rho'(t')=\rho(t), \qquad s'(t')=s(t), \qquad u'(t')=u(t).
\ee

As a sample calculation used in the derivation of (\ref{tr1}), we display below the chain of relations involving the infinitesimal parameter $\beta$
\bea
&&
e^{i\beta D} e^{i\rho(t)P}=e^{i\rho(t)P} e^{-i\rho(t)P} e^{i\beta D} e^{i\rho(t)P}=e^{i\rho(t)P} e^{-i\rho(t)P} \left(1+i\beta D\right) e^{i\rho(t)P}
\nonumber\\[4pt]
&&
\quad \quad \quad \quad ~  =e^{i\rho(t)P} \left(1+i\beta e^{-i\rho(t)P} D e^{i\rho(t)P}\right)=e^{i\rho(t)P} \left(1+i\beta[D+\rho(t)P]\right)
\nonumber\\[4pt]
&&
\quad \quad \quad \quad ~ =e^{i\rho(t)\left(1+\beta\right)P}e^{i\beta D}.
\eea
Note that the Baker--Campbell--Hausdorff formula was used on the last but one step only. Because $\beta$ is infinitesimal, one can approximate $\left(1+i\beta[D+\rho(t)P]\right)$ by
$e^{i\beta \rho(t) P}e^{i\beta D}$.

As the second step, one computes the Maurer--Cartan one--forms
\be
\tilde g^{-1} d \tilde g=i\left(\omega_H H+\omega_P P +\omega_K K+\omega_D D\right),
\ee
where
\be\label{inv}
\omega_H=dt, \qquad \omega_P=\dot\rho e^{-u} dt, \qquad \omega_K=e^u \left(\dot s+s^2 \dot\rho\right)dt, \qquad \omega_D=\left(\dot u-2 s \dot\rho\right) dt,
\ee
which are invariant under the transformation $\tilde g'= g \cdot \tilde g$ represented by Eqs. (\ref{tr1}), (\ref{tr2}) above. These
provide convenient building blocks for constructing invariant action functionals or equations of motion.

If, for some reason, it is desirable to reduce the number of the independent Goldstone fields, one can use (\ref{inv}) to impose constraints. For instance, choosing the following restrictions:
\be
\omega_P-\mu \omega_H=0, \qquad \omega_D+2 \nu \omega_H=0,
\ee
where $\mu$ and $\nu$ are arbitrary constants, one can express $u$ and $s$ in terms of $\rho$
\be\label{constr}
e^{-u}=\frac{\mu}{\dot\rho}, \qquad s=\frac{\nu}{\dot\rho}+\frac{\ddot\rho}{2 \dot\rho^2}~.
\ee
Substituting these relations into the remaining form $\omega_K$, multiplying by $2\mu$, and subtracting $2\nu^2 \omega_H$, one gets
\be
2\mu \omega_K-2\nu^2 \omega_H=S(\rho(t)) dt,
\ee
with $S(\rho(t))$ defined in (\ref{SD}).

Thus, the Schwarzian derivative arises quite naturally, if one applies the method of nonlinear realizations to $SL(2,R)\times R$ group and decides to keep the number of the independent Goldstone fields to a minimum. Note that within the group--theoretic framework the $SL(2,R)$--invariance of the derivative is obvious as it is built in terms of the invariant Maurer--Cartan one--forms.

\vspace{0.5cm}

\noindent
{\bf 3. Lagrangian formulation for the Schwarzian mechanics}\\

\noindent
In a recent work \cite{AG}, a variant of the Schwarzian mechanics was studied which was obtained by setting $S(\rho(t))$
to be equal to a coupling constant $\lambda$
\be\label{eom}
S(\rho(t))=\lambda.
\ee
The consideration in the preceding section allows us to immediately build a Lagrangian formulation which reproduces (\ref{eom}).

Consider the action functional composed of the invariant Maurer--Cartan one--forms (\ref{inv})
\be\label{act}
\int dt~ \omega_H^{-2} \left( \omega_P \omega_K+\nu \omega_H \omega_D\right)=\int dt ~\dot\rho\left(\dot s +s^2 \dot\rho-2 \nu s\right),
\ee
where $\nu$ is an arbitrary nonzero constant. Note that in (\ref{act}) we dropped the total derivative term $\nu\dot u$.
A variation of the action with respect to $s$ yields the equation
\be\label{cond1}
s=\frac{\nu}{\dot\rho}+\frac{\ddot\rho}{2 \dot\rho^2}~,
\ee
which links $s$ to $\rho$. This is identical to the rightmost constraint in Eq. (\ref{constr}) above.
Varying the action with respect to $\rho$, one finds
\be\label{cond2}
\frac{d}{dt} \left(\dot s +2 s^2 \dot\rho-2 \nu s \right)=0.
\ee
Upon substitution of (\ref{cond1}) into (\ref{cond2}), one gets
\be
\frac{d}{dt} \left(\dot s +2 s^2 \dot\rho-2 \nu s \right)=\frac{1}{2\dot\rho} \frac{d}{dt}S(\rho(t))=0 \quad \Rightarrow \quad S(\rho(t))=\lambda,
\ee
where $\lambda$ is a constant of integration. Identifying the latter with a coupling constant, one reproduces the dynamical system (\ref{eom}).

Because the action functional (\ref{act}) is given in terms of the Maurer--Cartan one--forms, it does not change under the transformations (\ref{tr1}) and (\ref{tr2}). Constructing the Noether charges by conventional means\footnote{
Lagrangian symmetries, which we consider in this work, are of the form
$t'=t+\delta t (t)$, $x'_i (t')=x_i(t)+\delta x_i(t,x(t))$.
If the action $S=\int dt \mathcal{L}(x,\dot x)$ holds invariant up to a total derivative,
$\delta S=\int dt \Big( \frac{d F}{dt}\Big)$,
the conserved quantity is derived from
$\delta x_i \frac{\partial \mathcal{L}}{\partial \dot x_i}-
\delta t \Big( \dot x_i \frac{\partial \mathcal{L}}{\partial \dot x_i}-\mathcal{L} \Big)-F$
by discarding an infinitesimal parameter of the transformation. For the case at hand, only the special conformal transformation associated with the last line in (\ref{tr1}) yields a total derivative term,  $F=-2\nu\gamma\rho$.}, one finds
\begin{align}\label{IM}
&
H=\dot\rho\left(\dot s+s^2 \dot\rho \right), && P=\dot s +2 s^2 \dot\rho-2 \nu s,
\nonumber\\[4pt]
&
D=\rho P-s \dot\rho, &&
K=\rho^2 P+(1-2s \rho)\dot\rho+2\nu\rho.
\end{align}
Taking into account the condition (\ref{cond1}) and the equation of motion (\ref{eom}), one finds that $H$ degenerates to a constant
\be
H=\frac 12 \lambda+\nu^2,
\ee
while $P$, $D$, and $K$ yield the integrals of motion associated with Eq. (\ref{eom})
\bea\label{IM1}
&&
P=\frac{1}{2\dot\rho}\left(\lambda+\frac 12 {\left(\frac{\ddot\rho}{\dot\rho}\right)}^2 \right), \qquad
D=\rho P-\frac{\ddot\rho}{2\dot\rho}, \qquad
K=\rho^2 P+\dot\rho-\frac{\rho\ddot\rho}{\dot\rho}
\eea
The expressions for $P$ and $D$ agree with those found in \cite{AG} by the explicit integration of (\ref{eom}), while $K$ proves to be functionally dependent
\be
P K-D^2-\frac{\lambda}{2}=0.
\ee

Concluding this section, it is worth emphasising that, within the context of the Schwarzian mechanics, $SL(2,R)$ group acts in the space of the Goldstone fields (\ref{tr1}) by affecting their form only. This is to be contrasted with the $1d$ conformal mechanics by de Alfaro, Fubini and Furlan \cite{DFF}, in which $SL(2,R)$ is realized in the $1d$ space parametrized by the temporal variable $t$ (cf. (\ref{tr1}))
\be
t'=t+\alpha+\beta t+\gamma t^2.
\ee
This is accompanied by the transformation law of the Goldstone field
\be
\rho'(t')=\rho(t)+\frac{1}{2} (\beta+2\gamma t) \rho(t)
\ee
and gives rise to the {\it second} order invariant equation
\be
\ddot\rho(t)=\frac{\lambda^2}{\rho(t)^3},
\ee
where $\lambda$ is a coupling constant. The derivation of the conformal mechanics \cite{DFF} by the method of nonlinear realizations was reported in \cite{IKL}.

\vspace{0.5cm}

\noindent
{\bf 4. Geometric description of the Schwarzian mechanics}\\

\noindent
Within the general relativistic framework, the conventional method of describing a classical mechanics systems with $d$ degrees of freedom is to embed its equations of motion into the null geodesic equations associated with a Brinkmann–-type metric defined on $(d+2)$--dimensional spacetime of the Lorentzian signature \cite{LE}. Let us discuss a geometric formulation for the Schwarzian mechanics (\ref{eom}).

As the first step, one constructs the Hamiltonian\footnote{Interestingly enough, although the Hamiltonian (\ref{h2d}) does describe a higher derivative theory, it is apparently not of the Ostrogradsky type.} corresponding to the action functional (\ref{act})
\be\label{h2d}
H_{2d}=p_s \left(p_\rho-s^2 p_s+2 \nu s \right),
\ee
where $(p_\rho,p_s)$ designate momenta canonically conjugate to the configuration space variables $(\rho,s)$. As the second step, one introduces two more canonical pairs $(t,p_t)$, $(v,p_v)$ and promotes $H_{2d}$ to the function which is homogeneously polynomial of degree two in the momenta (see Sect. 2 in \cite{GHKW} and related earlier work \cite{DGH})
\be
H_{2d} \quad \rightarrow \quad H_{4d}=p_s\left(p_\rho-s^2 p_s+2 \nu s p_v \right)+p_t p_v.
\ee
Identifying the latter with the $4d$ geodesic Hamiltonian $H_{4d}=\frac 12 g^{MN} p_M p_N$, in which $p_M=(p_t,p_v,p_\rho,p_s)$, one finally gets the Eisenhart metric
\be\label{EiM}
g_{MN} dz^M dz^N=2\left(dt dv-2 \nu s dt d\rho+s^2 d\rho^2+d\rho ds \right),
\ee
where we denoted $Z^M=(t,v,\rho,s)$. By construction, the null reduction of the geodesic equations associated with (\ref{EiM}) along $v$ reproduces (\ref{eom}).

A few comments are in order. Firstly, the metric (\ref{EiM}) is given in the global coordinate system. Secondly, it is of the ultrahyperbolic signature $(+,+,-,-)$, which is in agreement with the geometric analysis of higher derivative models in \cite{GM}. Thirdly, the metric admits five Killing vector fields
\bea
&&
\xi^M \partial_M=\partial_v, \qquad \qquad \quad \chi^M \partial_M=\partial_t, \qquad \qquad \quad  \phi^M \partial_M=\partial_\rho,
\nonumber\\[4pt]
&&
\psi^M \partial_M=\rho\partial_\rho-s\partial_s, \qquad
\zeta^M \partial_M=\rho^2 \partial_\rho+(1-2\rho s)\partial_s+2 \nu \rho \partial_v
\eea
of which $\partial_v$ is covariantly constant. Two of these can be used to construct the energy--momentum tensor
\bea
T_{MN}=-\frac{\nu^2}{4\pi} \xi_M \xi_N-\frac{1}{4\pi} (\xi_M \chi_N+\xi_N \chi_M),
\eea
which allows one to regard (\ref{EiM}) as a solution to the Einstein equations
\bea
R_{MN}-\frac 12 g_{MN} R=8\pi T_{MN}.
\eea
For more details concerning the Eisenhart lift of a generic $2d$ mechanics see a recent work \cite{FG}.

\vspace{0.5cm}

\noindent
{\bf 5. Conclusion}\\

\noindent
To summarize, in this work we applied the method of nonlinear realizations to $SL(2,R)\times R$ group and demonstrated that the Schwarzian derivative arose naturally, if one decided to keep the number of the independent Goldstone fields to a minimum. A Lagrangian formulation for a variant of the Schwarzian mechanics studied in \cite{AG} was constructed in terms of the invariant Maurer--Cartan one--forms and the full set of the integrals of motion was exposed. A geometric formulation has been constructed in terms of $4d$ metric of the ultrahyperbolic signature which obeys the Einstein equations.

Turning to possible further developments, it would be interesting to generalise the analysis in this work to the case of the super Schwarzian derivative (see, e.g., \cite{MTV} and references therein) and the supersymmetric Schwarzian mechanics. A possible link of the Hamiltonian formulation (\ref{h2d}) to the Ostrogradsky method is worth studying. Last but not least, it would be interesting to analyse if the method in Sect. 2 may result in other interesting higher order derivatives enjoying a given symmetry group.

\vspace{0.5cm}

\noindent{\bf Acknowledgements}\\

\noindent
This work was supported by the Russian Science Foundation, grant No 19-11-00005.


\end{document}